\documentclass[12pt]{article}
\usepackage{amsfonts}
\begin{document}
\thispagestyle{empty}
\newcommand{\be}{\begin{equation}}
\newcommand{\ee}{\end{equation}}
\newcommand{\sect}[1]{\setcounter{equation}{0}\section{#1}}
\newcommand{\vs}[1]{\rule[- #1 mm]{0mm}{#1 mm}}
\newcommand{\hs}[1]{\hspace{#1mm}}
\newcommand{\mb}[1]{\hs{5}\mbox{#1}\hs{5}}
\newcommand{\bea}{\begin{eqnarray}}
\newcommand{\eea}{\end{eqnarray}}
\newcommand{\wt}[1]{\widetilde{#1}}
\newcommand{\ux}[1]{\underline{#1}}
\newcommand{\ov}[1]{\overline{#1}}
\newcommand{\sm}[2]{\frac{\mbox{\footnotesize #1}\vs{-2}}
           {\vs{-2}\mbox{\footnotesize #2}}}
\newcommand{\prt}{\partial}
\newcommand{\eps}{\epsilon}\newcommand{\p}[1]{(\ref{#1})}
\newcommand{\R}{\mbox{\rule{0.2mm}{2.8mm}\hspace{-1.5mm} R}}
\newcommand{\Z}{Z\hspace{-2mm}Z}
\newcommand{\cd}{{\cal D}}
\newcommand{\cg}{{\cal G}}
\newcommand{\ck}{{\cal K}}
\newcommand{\cw}{{\cal W}}
\newcommand{\vj}{\vec{J}}
\newcommand{\vl}{\vec{\lambda}}
\newcommand{\vz}{\vec{\sigma}}
\newcommand{\vt}{\vec{\tau}}
\newcommand{\poiss}{\stackrel{\otimes}{,}}
\newcommand{\tx}{\theta_{12}}
\newcommand{\tb}{\overline{\theta}_{12}}
\newcommand{\zw}{{1\over z_{12}}}
\newcommand{\sqp}{{(1 + i\sqrt{3})\over 2}}
\newcommand{\sqm}{{(1 - i\sqrt{3})\over 2}}
\newcommand{\NP}[1]{Nucl.\ Phys.\ {\bf #1}}
\newcommand{\PLB}[1]{Phys.\ Lett.\ {B \bf #1}}
\newcommand{\PLA}[1]{Phys.\ Lett.\ {A \bf #1}}
\newcommand{\NC}[1]{Nuovo Cimento {\bf #1}}
\newcommand{\CMP}[1]{Commun.\ Math.\ Phys.\ {\bf #1}}
\newcommand{\PR}[1]{Phys.\ Rev.\ {\bf #1}}
\newcommand{\PRL}[1]{Phys.\ Rev.\ Lett.\ {\bf #1}}
\newcommand{\MPL}[1]{Mod.\ Phys.\ Lett.\ {\bf #1}}
\newcommand{\BLMS}[1]{Bull.\ London Math.\ Soc.\ {\bf #1}}
\newcommand{\IJMP}[1]{Int.\ J.\ Mod.\ Phys.\ {\bf #1}}
\newcommand{\JMP}[1]{Jour.\ Math.\ Phys.\ {\bf #1}}
\newcommand{\LMP}[1]{Lett.\ Math.\ Phys.\ {\bf #1}}
\newpage
\setcounter{page}{0} \thispagestyle{empty} \vs{12}

\renewcommand{\thefootnote}{\fnsymbol{footnote}}

\vspace{.5cm}
\begin{center}
{\large\bf Exceptional Structures in Mathematics and Physics and the Role of the Octonions} \vspace{.5cm}\\

\vs{10} {\large Francesco Toppan} ~\\ \quad
\\
 {\large{\em CBPF - CCP}}\\{\em Rua Dr. Xavier Sigaud
150, cep 22290-180 Rio de Janeiro (RJ)}\\{\em Brazil}\\

\end{center}
{\quad}\\ \centerline{ {\bf Abstract}}

\vs{6}

There is a growing interest in the logical possibility that exceptional
mathematical structures (exceptional Lie and superLie algebras, the exceptional
Jordan algebra, etc.) could be linked to an ultimate ``exceptional"
formulation for a Theory Of Everything (TOE).
The maximal division algebra of the octonions can be held as the mathematical
responsible for the existence of the exceptional structures mentioned above.
In this context it is quite motivating to systematically investigate the
properties of octonionic spinors and the octonionic 
realizations of supersymmetry. In particular the $M$-algebra can be consistently
defined  for two structures only, a real structure, leading to the standard $M$-algebra,
and an octonionic structure. The octonionic version of the $M$-algebra admits striking properties 
induced by octonionic $p$-forms identities. 

\section{Introduction.}
The search of an ultimate ``Theory Of Everything" (TOE) corresponds to a very 
ambitious and challenging program. At present the conjectured $M$-Theory (whose dynamics is yet to unravel),
a non-perturbative
theory underlying the web of dualities among the five consistent superstring theories and the maximal
eleven-dimensional supergravity, is regarded by perhaps the majority of physicists as the most promising
candidate for a TOE  \cite{tow}.\par
While a TOE can be minimally defined as a consistent theory providing the unified dynamics
of the known interactions, including the quantum gravity, a more ambitious viewpoint could be advocated, namely
that the TOE is {\em unique}. Features like the dimensionality of the space-time
should not be externally imposed to reproduce the known results, rather should arise naturally due to the consistency conditions of the theory.
It is quite pleasing that the superstring/$M$ theory program, at least at the Planck scale, satisfies 
this requirement.\par
In the last two or three decades several physicists and groups have investigated the possibility that, since the
universe is exceptional, it should be better grasped by an ``exceptional" ultimate TOE, see \cite{ram} and references therein. The word ``exceptional" is here
used technically, meaning an exceptional mathematical structure, such as the finite and sporadic Monster group,
the exceptional Lie and superLie algebras or the exceptional Jordan algebra (which will be briefly discussed
in the following).
\par
It would be highly desirable that miraculous features such as the hypercharge balance of the quarks-leptons in each of the three known families, leading to the cancellation of the chiral anomaly and the perturbative consistency of
the electroweak standard model, would be ``explained" by an underlying exceptional theory \cite{wit}.\par
This research program, despite being highly conjectural, has a strong merit. Focusing on 
exceptional mathematical structures (in contrast with the standard ``boring" ones) helps clarifying their role
and discovering their mutual, sometimes unsuspected, relations. In the following, several examples will be given.\par
For what concerns the physical side, various motivations  and scattered pieces of evidence for the proposed program can be given. We will list here three of them:
\par
{\em i}) At the level of $4D$ GUT, \cite{{ram},{wit}} the Georgi-Glashow $SU(5)$ model can be consistently embedded in the group chain
$SU(5)\subset SO(10)\subset E_6$, where the last group corresponds to an exceptional group (the ``exceptional" viewpoint can explain why $SU(5)$ appears instead of $SU(N)$ for an arbitrary value of $N$),
\par
{\em ii}) The Heterotic String is consistently formulated for the $E_8\times E_8$ group, which contains
as subgroups the above chain of GUT groups (please notice that $E_8$ is not only exceptional, it is the {\em maximal}
exceptional group),\par
{\em iii}) finally, the phenomenological requirement of dealing with a $4$-dimensional $N=1$ supersymmetric chiral
theory implies a Kaluza-Klein compactification from the eleven-dimensional $M$-theory, based on a seven-dimensional
singular manifold with $G_2$ holonomy \cite{duf}. Please notice that $G_2$ is, once more, an exceptional group. This $M$-theory/$11D$ SUGRA prescription should be compared with the
Calabi-Yau six-dimensional compactifications from perturbative strings admitting $SU(3)$ (a standard group) holonomy.\par
It is worth to recall very basic mathematical properties of some of the exceptional structures mentioned so far.
For what concerns simple Lie algebras, besides the four classical series $A_n$, $B_n$, $C_n$, $D_n$  associated to the
special unitary, special orthogonal and symplectic groups, five exceptional Lie algebras appear in the Cartan-Killing
classification: $G_2$, $F_4$, $E_6$, $E_7$ and $E_8$, the respective numbers denoting their fixed rank. We already encountered three of them, $E_6$, $E_8$ and $G_2$ in the discussion above.The maximal exceptional Lie algebra,
$E_8$, is in some sense the most important of all. All lower-rank Lie algebras, both exceptional and not, 
can be recovered from $E_8$ through the coupled procedure of taking subalgebras and eventually folding them
to produce the non simply-laced simple Lie algebras \cite{fol}. \par
If we extend our considerations to Lie superalgebras, just two exceptional Lie superalgebras, admitting no
free parameter, exist. They are denoted as $G(3)$ and $F(4)$ \cite{dic}.\par
It is quite remarkable that the existence of all these exceptional algebras and superalgebras is mathematically motivated by their construction through the maximal division algebra, the division algebra of the octonions.
The fixed rank of the exceptional Lie (super)algebras is a consequence of the non-associativity of the octonions.
In contrast, the classical series of the $A_n$, $B_n$, $C_n$, $D_n$ algebras are related to the associative division algebras of real, complex and quaternionic numbers. 
\par
Indeed, $G_2$ is the $14$-component Lie algebra of the group of automorphisms of the octonions \cite{bae}. The remaining bosonic exceptional
Lie algebras are recovered from the so-called Tits's magic square construction \cite{bs}, associating a Lie algebra to any given pair of alternative division algebras and Jordan algebras (over a division algebra). By taking the Jordan
algebra to be the exceptional Jordan algebra $J_3({\bf O})$ of $3\times 3$ octonionic-valued hermitian matrices,
$F_4$, $E_6$, $E_7$ and $E_8$ are recovered if the alternative division algebra is respectively set equal to
${\bf R}$, ${\bf C}$, ${\bf H}$ and ${\bf O}$. A supersymmetric construction based on the octonions also
exists for the superalgebras $G(3)$ and $F(4)$, see \cite{sud}. We can say that the octonions are at the very core of all these exceptional structures. \par
Incidentally, we have introduced the exceptional Jordan algebra $J_3({\bf O})$, which can be easily proven admitting $F_4$ as the Lie algebra of its group of automorphisms. $J_3({\bf O})$ plays a very peculiar and perhaps not yet fully understood
role which goes back to the early days of quantum mechanics.
The Jordan formulation of quantum mechanics \cite{jor}, based on the Jordan product between observables, is equivalent to the
standard formulation as far as associative structures are taken into account. It was realized by the fathers of quantum mechanics, Jordan himself, von Neumann and Wigner \cite{jnw}, that one and only one notable exception exists,
$J_3({\bf O})$, as algebra of observables. $F_4$ plays here the role of the unitary transformations (i.e. the dynamics),
while the states are associated to the non-desarguian Moufang octonionic projective plane
 ${\bf O}{\bf P}^2$ \cite{gpr}. So far no physical system has yet been found to be described by this unique octonionic quantum
 mechanical system. The octonionic quantum mechanics is, at present, just a consistent mathematical curiosity.
 Interestingly, however, $J_3({\bf O})$ has been recently proposed (by showing it possesses the correct properties) to define an
 exceptional matrix Chern-Simon theory \cite{smo} which could grasp, within the exceptional viewpoint discussed above, the features of a background-independent formulation for quantum gravity.
 \par
We are therefore naturally led from the ``exceptional considerations" to investigate the division algebra of the
octonions, see also \cite{boy}. There is another, partially independent, argument justifying the interest in the octonions.
Since the octonions are the maximal division algebra (real, complex and quaternionic numbers can be obtained as subalgebras), they can be regarded on the same footing as, let's say, the eleven dimensional maximal supergravity.
In a TOE viewpoint it makes sense, as already recalled, investigating the largest algebraic setting available,
without imposing any unnecessary external constraint. It is true that the non-associativity of the octonions
makes some issues problematics (it is e.g. unclear how to construct octonionic tensor products \cite{kr}). 
Similar issues cannot however be raised as strong objections towards a work which is necessarily still
in progress.\par
It is worth mentioning that in the seventies octonions have been investigated as possible algebraic explanations for the
absence of quarks and other colored states of the strong interactions \cite{gg}. Analogous ideas, from time to time, resurface
again. The unobservability of the extra dimensions (with no need of introducing a Kaluza-Klein compactification or
a brane-world scenario) by using an octonionic description, has been suggested as a possibility in \cite{md}.\par
Nowadays the arena where the use of octonions seems most promising concerns supersymmetry. This is of course based on the celebrated connection of supersymmetry with division algebras \cite{kt}. Octonionic supersymmetry is a combination of two remarkable ideas
and beautiful mathematical structures which, so far, have not yet been detected on physical experiments.\par
This large introduction was meant to motivate investigating the octonionic supersymmetry \cite{top}.
The rest of this paper consists in a quick introduction to the octonionic supersymmetry.
The most important result here discussed consists in the octonionic formulation of the $M$-algebra \cite{{lt1},{lt2}}, with its surprising properties derived from octonionic $p$-form identities \cite{crt1} 
leading to the equivalence of the octonionic
$5$-brane sector ($M5$) with the octonionic $1$ and $2$-branes sectors ($M1$ and $M2$), symbolically $M5\equiv M1+M2$.\par
The recognition of the importance of the octonionic $M$-algebra derives from the fact that an $11$-dimensional
$M$-algebra can be consistently defined using two and only two structures, i.e. the real (${\bf R}$) structure and the
octonionic (${\bf O}$) structure. While the real structure leads to the standard formulation of the $M$-algebra,
the octonionic structure is potentially linked with the exceptional structures discussed above.
If we combine together the two assumptions previously discussed, i.e. that
\par
{\em i}) the TOE is succesfully described by the $M$-theory and its related $M$-algebra and
\par
{\em ii}) the TOE is based on an exceptional mathematical structure,\\
then,
the octonionic $M$-algebra naturally arises as the potential underlying algebra.

\section{Octonionic spinors.}

The seven imaginary octonions $\tau_i$ satisfy the algebraic relation
\begin{eqnarray}
\tau_i\cdot \tau_j &=& -\delta_{ij} + C_{ijk} \tau_{k},
\label{octonrel}
\end{eqnarray}
for $i,j,k = 1,\cdots,7$, where $C_{ijk}$ are the totally antisymmetric
octonionic structure constants given by
\begin{eqnarray}
&C_{123}=C_{147}=C_{165}=C_{246}=C_{257}=C_{354}=C_{367}=1&
\label{strconst}
\end{eqnarray}
and vanishing otherwise. The product (\ref{octonrel}) is non-associative, but satisfies
the weaker condition of alternativity \cite{bae}.\par
Due to the antisymmetric property of $C_{ijk}$, the anticommutator between
imaginary octonions satisfy
\begin{eqnarray}
\tau_i\cdot \tau_j +\tau_j\cdot \tau_i &=& -2\delta_{ij},
\end{eqnarray}
providing an octonionic realization of the Euclidean Clifford algebra $C(0,7)$
which is of course not equivalent w.r.t. the standard associative realization \cite{top}.\par
Non-associative octonionic realizations of Clifford algebras are available in higher-dimensional
space-times as well. This can be easily proven by noticing that two iterative algorithms exist allowing
the lifting of a $D$-dimensional space-time realization of a Clifford algebra into a $(D+2)$-dimensional one. Indeed, if $\gamma_a$ denote a realization of the $C(p,q)$ 
Clifford algebra (with $p$ space-like
and $q$ time-like directions), then
\begin{eqnarray}
 \Gamma_r &\equiv& \left(
\begin{array}{cc}
  0& \gamma_a \\
  \gamma_a & 0
\end{array}\right), \quad \left( \begin{array}{cc}
  0 & {\bf 1}_d \\
  -{\bf 1}_d & 0
\end{array}\right),\quad \left( \begin{array}{cc}
  {\bf 1}_d & 0\\
  0 & -{\bf 1}_d
\end{array}\right)\nonumber\\
&&\nonumber\\ (p,q)&\mapsto&
 (p+1,q+1)\label{one}
\end{eqnarray}
and
\begin{eqnarray}
 \Gamma_r &\equiv& \left(
\begin{array}{cc}
  0& \gamma_a \\
  -\gamma_a & 0
\end{array}\right), \quad \left( \begin{array}{cc}
  0 & {\bf 1}_d \\
  {\bf 1}_d & 0
\end{array}\right),\quad \left( \begin{array}{cc}
  {\bf 1}_d & 0\\
  0 & -{\bf 1}_d
\end{array}\right)\nonumber\\
&&\nonumber\\ (p,q)&\mapsto&
 (q+2,p),\label{two}
\end{eqnarray}
for $r=1, \ldots ,D+2$, produce a Clifford representation of signature $(p+1,q+1)$ and $(q+2,p)$
respectively.\par
It turns out in particular that the Clifford algebra associated with the
eleven dimensional minkowskian space-time $(10,1)$ admits an octonionic realization in
terms of $4\times 4$ Gamma matrices with octonionic entries by applying twice the 
two algorithms above, e.g. by sending
$(0,7)\rightarrow (9,0)\rightarrow (10,1)$ (please notice that each time one of the two algorithms
above are applied the size of the new Gamma matrices is doubled).
\par
The $32$ real-component spinors entering the $11D$ supergravity or standard $M$-theory can therefore be replaced by $4$ octonionic-component spinors (the total number of real components still being $32=4\times 8$).\par
It should be noticed that in an octonionic realization the commutators $\Sigma_{rs}=\relax [\Gamma_r,\Gamma_s]$ are no longer the generators of the generalized Lorentz group $SO(p,q)$.
They correspond instead to the generators of the coset $SO(p,q)/G_2$, $G_2$ being, as already recalled, the $14$-dimensional group of automorphisms of the octonions. The particular coset associated to the $7$-dimensional Euclidean space described by imaginary octonions corresponds to the seven sphere $S^7$ \cite{lmh}.

\section{Octonionic supersymmetry}

We have seen in the previous section that one can give an octonionic description for both
Clifford algebras and spinors. Since these are the basic ingredients entering supersymmetry,
it is quite tempting to apply the previous construction to describe octonionic supersymmetry.\par
Indeed octonionic supersymmetry can be introduced  both at the level of one dimensional octonionic 
supersymmetric quantum mechanics \cite{{top},{crt2}} and at the level of higher-dimensional supersymmetry
(even generalized supersymmetries, as it is the case for the octonionic $M$-algebra discussed in the
next section).\par
For what concerns the $1D$ octonionic supersymmetric quantum mechanics the relation can be understod on the basis of the one-to-one correspondence between irreducible representations of $1D$
$N$-extended supersymmetries on one side and the special subclass of $D$-dimensional Clifford 
Gamma-matrices of Weyl type (namely the ones which are in block-antidiagonal form and can therefore
be promoted to be ``fermionic" matrices of a superalgebra). The correspondence is obtained by identifying $N$ (the number of extended supersymmetries) with $D$ (the dimensionality of the space-time), i.e. $N=D$ (see \cite{pt} for details). The octonionic counterpart of this construction has been discussed in \cite{{top},{crt2}}, while an application to an $N=8$ octonionic supersymmetric dynamical system has been introduced in \cite{crt3}  (b.t.w. $8$ is the minimal number
of $1D$ extended supersymmetries which can be ``octonionically" realized).\par
For what concerns the supersymmetry in higher dimension an interesting application regards the so-called
generalized space-time supersymmetries, namely the ones going beyond
the standard H{\L}S scheme \cite{hls}. This implies that the
bosonic sector of the Poincar\'e or conformal superalgebra no
longer can be expressed as the tensor product structure
$B_{geom}\oplus B_{int}$, where $B_{geom}$ describes space-time
Poincar\'e or conformal algebras and the remaining generators
spanning $B_{int}$ are Lorentz-scalars.\par In the particular case
of the Minkowskian $D=11$ dimensions, where the $M$-theory should
be found, in the standard case the following construction is allowed. 
As already recalled in this case the spinors are
real and have $32$ components.
By taking the anticommutator of two such spinors the most general
expected result consists of a $32\times 32$ symmetric matrix with
$32+\frac{32\cdot31}{2}=528$ components. On the other hand, the
standard supertranslation algebra underlying the maximal
supergravity contains only the $11$ bosonic Poincar\'e generators
and by no means the r.h.s. saturates the total number of $528$.
The extra generators that should be expected in the right hand
side are obtained by taking the totally antisymmetrized product of
$k$ Gamma matrices (the total number of such objects is given by
the Newton binomial ${\textstyle{\left(\begin{array}{c}
  D \\
  k
\end{array}\right)}}$).
Imposing on the most general $32\times 32$ matrix the further
requirement of being symmetric, the total number of $528$ is
obtained by summing the $k=1$, $k=2$ and $k=5$ sectors, so that
$528=11+55+462$. The most general supersymmetry algebra in $D=11$
can therefore be presented as
\begin{equation}
\{Q_a,Q_b\}= (A\Gamma_\mu )_{ab}P^\mu +(A\Gamma_{[\mu\nu]})_{ab}
Z^{[\mu\nu]} +
(A\Gamma_{[\mu_1\dots\mu_5]})_{ab}Z^{[\mu_1\dots\mu_5]}
\label{Malg}
\end{equation}
(where the real matrix $A$ is equivalent to $\Gamma_0$ \cite{crt1}).\par
$Z^{[\mu\nu]}$ and $Z^{[\mu_1\dots\mu_5]}$ are tensorial central
charges, of rank $2$ and $5$ respectively. These two extra central
terms on the right hand side correspond to extended objects
\cite{{bst},{dk}}, the $p$-branes. The algebra (\ref{Malg}) is
called the $M$-algebra. It provides the generalization of the
ordinary supersymmetry algebra recovered by setting $Z^{[\mu\nu]}
\equiv Z^{[\mu_1\dots\mu_5]}\equiv 0$.
It is this construction that we are going to analyze in the next
section in the case of the octonionic structure, namely for minkowskian
eleven-dimensional spinors described by $4$ octonionic components.

\section{The octonionic $M$-algebra.}

The octonionic counterpart of the eleven dimensional $M$ algebra (\ref{Malg}) is given
by the expression 
\begin{eqnarray}
\{ Q_a, {Q_b}^\ast \} &=& {\cal Z}_{ab},\label{gensusyalg}
\end{eqnarray}
where ${Q_a}^\ast$ denotes the principal conjugation in the given
division algebra \cite{bae}. The spinors are octonionic $4$-component and the r.h.s.
is an octonionic-valued $4\times 4$ hermitian matrix
 ${\cal Z}_{ab} =
{\cal Z}_{ba}^\ast$, whose total number of independent components, in the real counting,
is given by $52= 4+\frac{4\times 3}{2}\times 8$.\par
Just as the real case, the r.h.s. can be expanded in the antisymmetric product of
octonionic-valued Gamma-matrices. The number $52$, replacing the total number of $528$
bosonic components of the real case, represents the maximal number of saturated bosonic
components for the octonionic $M$ algebra. In order to understand how this number can be produced
and which are the octonionic $p$-forms entering the r.h.s., let's start with a brief digression concerning the
antisymmetrized product of octonionic Gamma matrices. Since the octonionic algebra is
non-associative a careful prescription has to be taken in order to correctly define such a product. Remarkably, this prescription turns out to be unique, if two requirements are chosen to
be satisfied:\par
{\em i}) that the octonionic $p$-forms fulfil the Hodge duality and
\par
{\em ii}) that the octonionic $p$-forms admit a well-defined character with respect to the hermiticity condition, being for a given $p$ either {\em all} hermitian or {\em all} antihermitian.\par
It can be easily shown that in an octonionic Gamma matrices realization of a $D$ dimensional Clifford algebra, out of the $D$ Gamma matrices, $D-7$ are purely real, while the remaining $7$
are proportional to the imaginary octonions.
It is therefore sufficient to correctly define the antisymmetrized products among $p$ imaginary
octonions. Since up to $p=2$ the associativity is guaranteed, the only case that we need to explicitly define corresponds to $p=3$ (the case $p=4$, since $4= 7-3$, automatically corresponds
to the Hodge dual case; similarly $p=5,6,7$ are Hodge-dual of $p=2,1,0$ respectively).
Due to the properties of imaginary octonions, only two inequivalent cases have to be examined,
either the three imaginary octonions are chosen to produce the associative  $su(2)$ subalgebra or not. In the first case, with the ({\ref{strconst}) choice of the octonionic structure constant, we can take
the three $\tau_i$ being given by $i=1,2,3$. Their antisymmetric product is proportional to the identity. The second case is non-associative and can be examined by just
taking $i=1,2,4$. In order to produce in this case as well an antisymmetric product 
$[\tau_{1}\cdot \tau_{2}\cdot  \tau_4]$
proportional to the identity, there is only one prescription which we are forced to take, namely
\begin{eqnarray}
\relax [\tau_{1}\cdot \tau_{2}\cdot  \tau_4]
&\equiv& 
\frac{1}{3!}
\sum_{perm.} (-1)^{\epsilon_{i_1 i_2 i_3}}
\frac{1}{2}((\tau_{i_1}\cdot \tau_{i_2}) \cdot \tau_{i_3}+ \tau_{i_1}\cdot(\tau_{i_2}\cdot
\tau_{i_3})),
\label{antisym}
\end{eqnarray}
for $i_j =1,2,4$.
The general formula for the antisymmetrized product of $p$ octonionic Gamma matrices is given in
\cite{crt1}. It should be noticed that the number of (real) components for octonionic $p$-forms
is reduced w.r.t. the associative (real, complex or quaternionic) cases. This is true already for $p=2$. This case can be easily understood since $14$ components are killed by the $G_2$ automorphisms of the octonions (we have, e.g. $[\tau_i ,\tau_j] = 2 C_{ijk}\tau_k$).\par
In $D$ odd dimensional octonionic spacetimes we get the following table, whose columns are
labeled by the antisymmetric product of Gamma matrices of rank $p$ and the entries denote
the total number of their components (real counting) 
{{\begin{eqnarray}&
\begin{tabular}{|c|c|c|c|c|c|c|c|c|c|c|c|c|c|c|}\hline
&$0$&$1$&$2$&$3$&$4$&$5$&$6$&$7$&$8$&$9$&$10$&$11$&$12$&$13$\\
\hline $D=7$&${\underline {1}}$&$7$&$7$&${\underline
{1}}$&${\underline{ 1}}$&$7$&$7$&${\underline{ 1}}$&&&&&&\\ \hline

$D=9$&${\underline {1}}$&${\underline {9}}$&$22$&$22$&$
{\underline {10}}$&${\underline {10}}$&$22$&$22$&${\underline
{9}}$&${\underline {1}}$&&&&\\ \hline

$D=11$&$1$&${\underline {11}}$&${\underline
{41}}$&$75$&$76$&${\underline {52}}$&${\underline
{52}}$&$76$&$75$&${\underline {41}}$&${\underline {11}}$&$1$&&\\
\hline

$D=13$&$1$&$13$&${\underline {64}}$&${\underline{
168}}$&$267$&$279$&${\underline{ 232}}$&${\underline{
232}}$&$279$&$267$&${\underline{ 168}}$&${\underline{
64}}$&$13$&1\\ \hline

\end{tabular}&\nonumber\end{eqnarray}}}
\begin{eqnarray}
&&\label{tableofgammas}
\end{eqnarray}
The hermitian components are underlined.\par

Identities relating
higher-rank antisymmetric octonionic tensors are expressed in the above table. Let us discuss 
here the $D=11$ case, relevant for the octonionic $M$ algebra.
The $52$ independent components of an
octonionic hermitian $(4\times 4)$ matrix can be expressed either
as a rank-$5$ antisymmetric tensors (simbolically denoted as
``$M5$"), or as the combination of the $11$ rank-$1$ ($M1$) and
the $41$ rank-$2$ ($M2$) tensors. The relation between $M1+M2$ and
$M5$ can be made explicit as follows. The $11$ vectorial indices
$\mu$ are split into $4$ real indices, labeled by $a,b,c,\ldots$
and $7$ octonionic indices labeled by $i,j,k,\ldots$. We get, on
one side, {{\begin{eqnarray}&
\begin{tabular}{cc}

$4$& $M1_a$\\

$7$&$M1_i$\\

$6$&$M2_{[ab]}$\\

$4\times 7= 28$&$M2_{[ai]}$\\

$7$& $ M2_{[ij]}\equiv M2_{i}$

\end{tabular}&\nonumber\end{eqnarray}}}

while, on the other side, {{\begin{eqnarray}&
\begin{tabular}{cc}

$7$& $M5_{[abcdi]} \equiv M5_i$\\

$4\times 7=28$&$M5_{[abcij]}\equiv M5_{[ai]}$\\

$6$&$M5_{[abijk]}\equiv M5_{[ab]}$\\

$4$&$M5_{[aijkl]}\equiv {M5}_a$\\

$7$& $ M5_{[ijklm]}\equiv {\widetilde M5}_{i}$

\end{tabular}&\nonumber\end{eqnarray}}}
which shows the equivalence of the two sectors, as far as the
tensorial properties are concerned. The correct
total number of $52$ independent components is recovered
\begin{eqnarray}
52 &=& 2\times 7 +28+6+4.
\end{eqnarray}
Please notice that the table (\ref{tableofgammas}) refers to the antisymmetric product
of Gamma matrices, which explains why $M5_i$ and ${\widetilde M5}_i$ are actually different. 
The octonionic equivalence of different antisymmetric tensors can
be symbolically expressed, in odd space-time dimensions, through
{{\begin{eqnarray}&
\begin{tabular}{|c|c|}\hline

$D=7$& $M0\equiv M3$\\ \hline

$D=9$&$M0+M1\equiv M4$\\ \hline

$D=11$&$M1+M2\equiv M5$\\ \hline

$D=13$&$M2+M3\equiv M6$\\ \hline

$D=15$&$M3+M4\equiv M0+M7$\\ \hline

\end{tabular}&\label{tablem}\end{eqnarray}}}

The octonionic $M$ algebra defined by (\ref{gensusyalg}) can therefore
be described 
either through the $11+41$ bosonic generators entering
\begin{equation}\label{eq1}
  {\cal{Z}}_{ab} = P^\mu (A\Gamma^{}_\mu)_{ab} +
   Z^{\mu\nu}_{\bf{O}} (A\Gamma^{}_{\mu\nu})_{ab}
   ,
\end{equation}
or through the $52$ bosonic generators entering
\begin{equation}\label{eq2}
  {\cal{Z}}_{ab} =
    Z^{[\mu_1\ldots \mu_5]}_{\bf{O}}
    (A\Gamma^{}_{\mu_1 \ldots
    \mu_5})_{ab}\, .
\end{equation}
Differently from the real case, the sectors specified by
(\ref{eq1}) and (\ref{eq2}) are not independent\cite{lt1}, leading
to an unexpected and far from trivial new structure in the
octonionic $M$-algebra.

\section{Conclusions.}

In this work we have discussed an approach to the ``Theory Of Everything" based on the
assumption that it could be described by an exceptional mathematical structure.
We have furnished scattered pieces of evidence of the arising of exceptional Lie algebras
and groups in the program of the unification of the interactions. We have further mentioned
that behind such exceptional structures one can find the division algebra of the octonions.
This is the maximal division algebra and in this respect plays a role similar to that of
maximal supergravity. In the technical part of this paper we have shown how to link octonions
with supersymmetry. We have in particular discussed the fact that the $M$ algebra, which is expected underlying
the $M$ theory, can arise in two and only two versions. One is based on real numbers and leads to
the standard description of $M$ algebra. The second one, however, is octonionic. As such,
it is potentially linked with the ``exceptional program" mentioned above. In formula (\ref{gensusyalg})
we presented the octonionic generalized supertranslation algebra in the minkowskian $D=11$
spacetime.
It is worth mentioning that this algebra admits a superconformal extension \cite{lt2} which
replaces the superalgebra $Osp(1,64|{\bf R})$ associated to the standard $M$ theory, with the
octonionic-valued superconformal algebra $Osp(1,8|{\bf O})$ associated to the octonionic $M$ algebra.
Striking new properties in the octonionic case, like the equivalence of different brane sectors,
have been thoroughly discussed.

\end{document}